\documentclass[11pt]{article}
\usepackage{vietnam,epsfig}

\def\Journal#1#2#3#4{{#1} {\bf #2}, #3 (#4)}

\def\NIMA{{\em Nucl. Instrum. Methods} A}
\def\NPB{{\em Nucl. Phys.} B}

\def\PRL{\em Phys. Rev. Lett.}

\def\ApP{{\em Astropart. Phys.}}
\def\MNRAS{{\em MNRAS}}

\def\AA{{\em Astronomy and Astrophysics}}

\def\be{\begin{equation}}
\def\ee{\end{equation}}
\def\bea{\begin{eqnarray}}
\def\eea{\end{eqnarray}}

\begin{document}
\vspace*{4cm}
\title{COSMIC-RAYS: AN UNSOLVED MYSTERY AT ALL ENERGIES!}

\author{ETIENNE PARIZOT }

\address{Institut de Physique Nucl\'eaire d'Orsay \\ IN2P3-CNRS/Universit\'e Paris-Sud, 91406 Orsay Cedex, France}

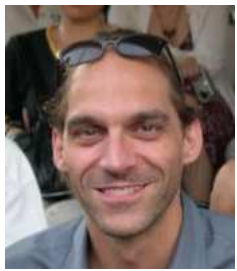
\maketitle\abstracts{We consider the phenomenology of cosmic-rays (CRs) and stress the interest of jointly studying their properties over the whole energy spectrum. While UHECRs are known to raise important physical and astrophysical problems, we recall that low-energy CRs also remain poorly understood, and we indicate the possibly important role of superbubbles in accelerating CRs up to a few EeV. We also investigate the viability of holistic models, and show that a unique type of sources producing CRs with a spectrum in $E^{-2.3}$ could in principle account for all the CRs in the universe.}

\section{Introduction}

Ultra-high energy cosmic rays (UHECRs) are known to challenge both observers and theoreticians, as they offer a few stimulating surprises, paradoxes or even mysteries (see other contributions in this volume). But this does not hold only for UHECRs! It has to be realised that a hundred years after the discovery of cosmic-rays (CRs), their sources -- even at low energy -- are still unknown. History and observational techniques have divided the CR spectrum into pieces: low-energy CRs, solar-modulated CRs, CRs in the GeV range, TeV range, accessible to satellite, balloon-borne or ground-based experiments, in the region of the \emph{knee}, around the \emph{ankle}, in the GZK range, up to ``super-GZK'' CRs. Yet, the only astrophysically motivated distinction is between Galactic cosmic-rays (GCRs) and extragalactic ones (EGCRs), and we shall make no other in this paper. Concerning the origin of GCRs, we shall briefly recall some problems of the standard model and the possibly central role played by superbubbles. We shall then ask ourselves in which energy range the transition from GCRs to EGCRs occurs, and finally, whether these two components could actually be but a single one, i.e. have the same astrophysical origin.

\section{On the sources of Galactic cosmic-rays}

Cosmic rays give rise to a number of \emph{id\'ees re\c cues} among most astrophysicists and high-energy physicists. It is generally believed that GCRs are well-known up to an energy of $\sim 10^{15}$~eV, that they are accelerated by the diffusive shock acceleration mechanism, and that this takes place at the expanding shocks of the intensively studied supernova remnants. This ``standard model'', however, does not work well at all!\cite{Parizot+01,Parizot+04,Parizot+05} At the high energy end of the spectrum, ``common knowledge'' holds it that CRs above $10^{18}$--$10^{19}$~eV are a deep mystery, that they are extragalactic, with unknown sources, and that their spectrum extends up to $3\,10^{20}$~eV without the expected \emph{GZK cut-off} (due to UHECRs interactions with the CMB\cite{Greisen66,ZatKuz66}). The situation, however, is not so clear-cut. There are actually several models able to account for the acceleration of UHECRs, and the violation of the GZK effect is not firmly established, even with the AGASA data set.\cite{DeMarco+03} Besides, it is clear what kind of spectrum should really be expected at the highest energies, depending on the source distribution, the extragalactic magnetic field,\cite{Deligny+04} the source spectrum and composition, etc. Finally, at intermediate energies, the nature of the spectral feature called the \emph{knee} (at $3\,10^{15}$~eV) is still uncertain, as is that of the \emph{ankle} ($\sim 3\,10^{18}$~eV), and other important, though less popular, questions deserve in-depth consideration, such as whether there is a second knee around $5\,10^{17}$~eV and what its nature is, or why there is no reported structure in the spectrum at low energy (e.g. below the knee), where local sources and/or structures in the CR propagation medium would be expected to leave some traces.

As can be seen, CRs raise a lot of fundamental questions, not only at ultra-high energy, but over the whole spectrum. The main reason why the CR sources remain unknown is that the charged particle trajectories are bent in the Galactic magnetic fields, and CRs do not point back to their sources. Their transport is essentially diffusive, up to an energy of a few $10^{18}$~eV, and indeed no clear indication of a meaningful anisotropy in the CR angular distribution has been obtained. Likewise, their energy spectrum cannot be easily used to reveal the sources, as a power-law, by essence, does not contain much information, apart from its logarithmic slope. In addition, the observed slope (roughly -2.7 below the knee, and -3.0 above it) is related to the source power law index through a (supposedly) power-law energy dependence of the CR confinement time in the Galaxy, whose index is also uncertain. According to the standard model, the source spectrum would be in $E^{-x}$, with $x = 2.0$--2.1, implying a confinement time in $E^{-\alpha}$, with $\alpha = 0.6$--0.7. However, in the most natural scenario of CR transport in the Galaxy, the energy dependence of the confinement time is related to a diffusion coefficient in $E^\alpha$, and theoretical considerations favour values of $\alpha\simeq 1/3$, so that $x \sim 2.3$--2.4 (cf. Eq.~(\ref{eq:lowEDensity}) below). The source spectrum itself is thus quite uncertain.

The other observable which could help determining the source of GCRs is the composition. But the latter does not provide much information either, as it is found to be essentially compatible with the standard interstellar medium (ISM), so that any mechanism accelerating particles out of an average ambient gas would qualify. An important exception is provided by the $^{22}$Ne abundance, which is much higher among GCRs than in the ISM.\cite{Meynet+01} Since this isotope is produced by very massive stars, it is argued that CRs originate from (or are contaminated by) regions where massive stars are concentrated. Another strong argument in favour of such an association is provided by the study of light element nucleosynthesis. Li, Be and B nuclei have been continuously produced in the Galaxy, over billions of years, by the interaction of CRs with the ISM, through spallation reactions implying C, N and O nuclei. In the 90's, it has been understood that the Li, Be and B nucleosynthesis cannot be accounted for by the standard model of CR origin, and that a natural solution could be obtained within the so-called \emph{superbubble model},\cite{Parizot00} in which most CRs are produced in superbubbles (i.e. large Galactic structures blown by the joint activity of many massive stars and the explosion of tens of supernov\ae), by collective acceleration mechanisms.\cite{Parizot+04,Parizot+05} Interestingly enough, such a scenario also accounts very naturally for the abnormal abundance of $^{22}$Ne among GCRs:\cite{Parizot+05} the fraction of $^{22}$Ne-rich massive stars ejecta which is required to explain the CR abundances is of the order of 3--6\%,\cite{Meynet+01} which is the same as that required to solve the problems of light element nucleosynthesis,\cite{Parizot00} and also the same as that expected from theoretical considerations of superbubble evolution.\cite{ParDru99}

Another very important problem of the standard model for CR origin is the maximum energy which protons can reach at the shocks of isolated supernov\ae~(SNe). It has been known for more than three decades that diffusive shock acceleration can hardly push protons up to energies higher than a few $10^{14}$~eV,\cite{LagCes83} i.e. below the knee of the spectrum. Recent developments have raised hopes that energies as high as $Z\times 10^{16}$~eV could be reached (although without observational support yet), but such an energy remains much too low to account for GCRs in the ankle energy range, even if these are essentially Fe nuclei. Now it is extremely improbable that the main source of GCRs (isolated SNe, according to the standard model) ends up around the knee (or $Z$ times higher for heavier nuclei) and a new -- currently unknown -- source of CRs takes over from a few $10^{16}$~eV up to the ankle, with an apparently perfect match of both the energy and the flux! If such an additional source of CRs is to be invoked to fill the gap between SN-accelerated CRs and CRs at the ankle, it might as well (and actually more easily) be invoked as the source of \emph{all} GCRs. In this respect, it should be noted that the above-mentioned superbubble model has offered a reasonable hope that cosmic-rays could be generated consistently over the whole spectrum, up to $Z\times 10^{17}$~eV (i.e. up to the ankle for Fe nuclei).\cite{BykTop01,Parizot+04} It should also be added that since most SNe are expected to explode inside superbubbles, the energetically motivated connection between SNe and GCRs does actually point towards superbubbles (rather than isolated SNe) playing a major role in the acceleration of CRs. Although much work is still needed to elucidate some aspects of such a model, it seems to be able to solve many of the long-standing problems of the standard scenario for the origin of GCRs. This is addressed in detail elsewhere.\cite{Parizot+01,Parizot+04,Parizot+05}

\section{Transition to extragalactic cosmic-rays (EGCRs)}

A crucial information for the global understanding of CRs is the energy range over which the transition from a Galactic component of CRs to an extragalactic one occurs. It has long be thought that the ankle is the natural feature in the CR spectrum where such a transition shows up. The softening of the slope (i.e. hardening) of the spectrum is indeed easy to interpret as the eventual taking over of a harder CR component which was subdominant at lower energy. It should be mentioned, however, that a different conclusion has recently been proposed in the wake of composition results presented by the HiRes Collaboration, tentatively showing a transition from light to heavy primary nuclei at an energy around $5\,10^{17}$~eV,\cite{ArcSok03,Thompson05} which could be identified with a (still to be confirmed) \emph{second knee} feature in the spectrum. Although a simple phenomenological model taking such an early G/EG transition into account has been proposed recently,\cite{Berezinsky+04} it still suffers from the necessary ``fine tuning'' accompanying any model of transition from a harder to a softer component. Nevertheless, it is worth recalling here that the location of the G/EG transition in the CR energy spectrum is not yet established firmly. This is an important blot in the global picture of CR phenomenology, and we suggest that this question should be the subject of intense observational and theoretical studies in the coming years.

From the observational point of view, two roads were followed to pinpoint the GCR/EGCR transition energy (range): one relying on the spectrum (change of slope) and the other on the composition (heavy to light transition). We suggest here that anisotropy studies might provide a more powerful tool. Indeed, it is expected that the Galactic component will become more and more anisotropic as the energy increases and the CR confinement by the Galactic magnetic fields becomes less effective. On the other hand, the incoming flux of extragalactic CRs should be more isotropic in the EeV range, even if the extragalactic fields are low, because of the large distance of the putative sources, allowing a diffusive transport regime to develop. While composition measurements prove very difficult in this energy range (they are indirect and depend on shower development models), the measurement of the CR anisotropy as a function of energy should be much less ambiguous, and help to explore this very important region of the spectrum.

\section{On the possibility of holistic cosmic-ray models}

Unless they consist of very heavy nuclei, with charges of the order of 100, their is no doubt that the highest energy CRs (with energies above $10^{20}$~eV) are produced outside the Galaxy (or at least outside the disk), because they cannot be confined by Galactic magnetic fields and their arrival direction pattern does not reflect the Galactic structure. The above distinction between a Galactic and an extragalactic CR component is thus unavoidable. However, while having a different spatial origin, the EGCRs and the GCRs could still have the same astrophysical origin. This is the essence of what could be called a \emph{holistic model} for CRs, i.e. a model in which \emph{all} the CRs, from the lowest to the highest energies, are produced in the same type of source. While such a requirement may seem to be an unnecessary a priori constraint on CR source models, one might conversely invoke Ockham's razor and claim that, since neither the sources of the GCRs nor those of the EGCRs are known, it may be simpler to assume that there is one type of CR source to be identified in the universe instead of two: \emph{pluralitas non est ponenda sine necessitate}...

Such a consideration is quite abstract, of course, since we are dealing with unknown or at least unspecified mechanisms, but it is nevertheless interesting to investigate whether holistic models are possible in principle -- from the phenomenological point of view.

As a matter of fact, if holistic sources of CRs existed in all galaxies like ours, the highest energy CRs would not be confined in the disks and halos, and would thus propagate throughout the universe (and fill it entirely if there is enough time for that), while the lower-energy ones (say up to the ankle -- or an appropriately scaled ankle depending on the size and magnetic field of each individual galaxy) would be confined for some time, and thus increase their density. Let us assume that all the CRs in the universe are (dominantly) produced by one type of sources, located inside galaxies similar to ours, with a power-law spectrum in $E^{-x}$. For the present purpose, it is equivalent to assume that CRs are produced homogeneously inside galaxies, with an differential injection rate, in $\mathrm{cm}^{-3}\mathrm{s}^{-1}\mathrm{eV}^{-1}$, given by:
\begin{equation}
q(E,\mathbf{r}) = q_{0}\left(\frac{E}{E_{0}}\right)^{-x}\times f(\mathbf{r}),
\label{eq:inputSpectrum}
\end{equation}
where $q_{0}$ is a normalisation factor (related to the global source power), $f(\mathbf{r}) = 1$ inside galaxies, and $0$ elsewhere.

In a first approximate approach, one may also assume that CRs of energy $E$ are confined inside galaxies within a volume $V_{\mathrm{conf}}$ and for a time $\tau_{\mathrm{conf}}(E)$, up to an energy $E_{\mathrm{c}}$ above which the CRs leak out immediately (more details will be found in a forthcoming paper). The total injection rate of CRs inside a galaxy,  per unit energy interval (in $\mathrm{s}^{-1}\mathrm{eV}^{-1}$), is then obtained by integration over the confinement volume:
\begin{equation}
\dot{N}(E) = \int_{V_{\mathrm{conf}}} q(E,\mathbf{r}) \mathrm{d}^3\mathbf{r} = q_{0}\left(\frac{E}{E_{0}}\right)^{-x}V_{\mathrm{conf}},
\label{eq:injectionRate}
\end{equation}
from which the density of low-energy CRs  (i.e. far below $E_{\mathrm{c}}$) inside a galaxy follows:
\begin{equation}
n_{\mathrm{G}}(E) = \frac{\dot{N}(E)\tau_{\mathrm{conf}}(E)}{V_{\mathrm{conf}}} = q_{0}\left(\frac{E}{E_{0}}\right)^{-x}\tau_{\mathrm{conf}}(E).
\label{eq:lowEDensity}
\end{equation}

Now the extragalactic density of high-energy CRs (i.e. above $E_{\mathrm{c}}$) is given by:
\begin{equation}
n_{\mathrm{EG}}(E) = \dot{N}(E)\times n_{\mathrm{gal}}\times T,
\label{eq:highEDensity}
\end{equation}
where $T$ is the ``age'' of the sources, i.e. the time since when they are active, and $n_{\mathrm{gal}}$ is the number of source galaxies per unit volume. This allows us to write a relation between the galactic and extragalactic CR densities:
\begin{equation}
n_{\mathrm{EG}}(E) = n_{\mathrm{G}}(E_{0}) \left(\frac{E}{E_{0}}\right)^{-x} \times V_{\mathrm{conf}}\times n_{\mathrm{gal}} \times \frac{T}{\tau_{\mathrm{conf}}(E_{0})}
\label{eq:nEG}
\end{equation}

In the above equation, the index, $x$, of the power-law source spectrum is the only unknown parameter. The CR density inside the Galaxy is obtained by direct measurement, say at $E_{0} = 10^{9}$~eV, just like the EGCR density at, say, $E = 10^{19}$~eV (which, incidentally, is the same inside and outside the Galaxy). One finds $n_{\mathrm{EG}}(10^{19}\,\mathrm{eV})/n_{\mathrm{G}}(1\,\mathrm{GeV}) \simeq 8\,10^{-29}$. Now the confinement time at $10^{9}$~eV is evaluated from cosmic-ray clock abundance data, within standard CR transport theory: $\tau_{0} \sim 2\,10^{7}$~yr. The age of the CR sources, $T$, is of the order of $10^{10}$~yr. The confinement volume is not very well-known for our Galaxy, as it depends on assumptions concerning the size of the magnetic field halo, but current propagation models favour halo heights of the order of 5~kpc (above and below the Galactic plane), so that $V_{\mathrm{conf}}\simeq \pi\times (15\,\mathrm{kpc})^2\times 10\,\mathrm{kpc}\simeq 7\,10^{-6}\,\mathrm{Mpc}^3$. Finally, the density of galaxies similar to ours can be evaluated from the 2dF Galaxy Redshift Survey:\cite{Norberg+02} $n_{\mathrm{gal}}\simeq 3\,10^{-3}\,\mathrm{Mpc}^{-3}$.\cite{LoeWax02}

Inserting the above numerical values into~Eq.~(\ref{eq:nEG}), one can solve for $x$ and obtain the slope of the power-law source spectrum which is required to make a holistic CR source model viable:
\begin{equation}
x \simeq 2.3.
\label{eq:x}
\end{equation}

Interestingly enough, such a value does not seem unreasonable at all. It is quite remarkable, actually, that not only a solution exists, making it possible that the same type of sources produce the CRs over the whole energy spectrum, but in addition this solution appears to be in keeping with the most natural expectations, both from theory (since the required slope is similar to that obtained from relativistic shock acceleration, or for internal shock models of GRBs, for instance) and from observations (since the UHECR data are very well fitted with a source spectrum in $E^{-2.3}$, provided that the CR source composition remains the same at all energies)\cite{Allard04}. It should also be reminded that this spectral index is also in perfect agreement with the low-energy CR phenomenology.

Obviously, the above considerations are no strong arguments in favour of holistic CR source models. But they show that such models seem to be possible, and in view of the general difficulty of understanding the CR phenomenology, not only at the high-energy end of the spectrum but also with respect to the more numerous and intensively studied Galactic CRs, they highlight the interest which one may find in considering the problem globally, in an approach including both GCRs and EGCRs either as two independent components, with particular attention to the transition region, or as a single component to be constrained further by forthcoming observational and phenomenological studies.

\section{Conclusion}

We have briefly recalled that, contrary to what is often believed, the Galactic cosmic-rays are not well understood, even at energies below the knee ($\sim 3\,10^{15}$~eV). We also mentioned the possible role of large Galactic structures known as superbubbles, which are the scene of strong massive star's activity (including winds and supernova explosions) and subsequent collective acceleration effects. These superbubbles may be the source of most of the GCRs, up to energies of the order of $10^{17}$~eV for protons, and up to the ankle for Fe nuclei.

On the other hand, UHECRs may not be such a complete mystery as often advertised. The main problem in UHECR phenomenology remains the putative absence of a GZK feature (strong suppression of the flux) above $10^{20}$~eV, but this fact remains controversial and, given the apparent discrepancy between the current observational results and the poor available statistics, it seems wise to wait for the results of the Pierre Auger Observatory, expected in the coming years, to conclude on this important matter. As far as UHECR acceleration is concerned, it does not seem so unreasonable that very energetic astrophysical sources such as gamma-ray bursts or active galactic nuclei (e.g. through their jets or hot spot shocks) be able to produce particles of a few $10^{20}$~eV. It is fair to say, however, that there is currently no preferred scenario for these UHECRs, and that the question of a non-standard physics origin is still open.

Another important unknown of the global CR phenomenology is the energy scale of the Galactic-extragalactic transition. Even if it occurs at the ankle (which remains the most natural energy scale for this transition), it will be quite instructive to study in greater detail its dynamics. We have indicated that anisotropy studies might be very useful in this energy range, as a supplement to the more difficult and ambiguous composition studies.

For all these reasons, we argue that it is important to address the problem of CRs in the universe through a global approach, not separating \emph{a priori} the phenomenon into two distinct problems (at respectively low and high energy). It may even be that the CRs at all energies come from one single type of sources, in which case the joint study of GCRs and EGCRs should provide efficient ways to constrain the models. As an illustration, we have shown that holistic CR source models are indeed possible, in principle, and that the required power-law source spectrum, in $E^{-2.3}$ or so, is in keeping with the main theoretical ideas about CR acceleration.

\section*{Acknowledgments}
I wish to thank the organisers for their invitation, as well as Andrei Bykov and Alexandre Marcowith for a long-time collaboration on the study of superbubbles and their link to cosmic-rays.

\end{document}